\documentclass[fleqn,usenatbib]{mnras}

\usepackage{newtxtext,newtxmath}

\usepackage[T1]{fontenc}
\usepackage{ae,aecompl}

\usepackage{graphicx}	
\usepackage{amsmath}	
\usepackage{amssymb}	
\usepackage{multirow}

\newcommand{\D}{\text{d}}
\newcommand{\x}{     \textbf{\textit{x}}}

\newcommand{\n}{\hat{\textbf{\textit{n}}}}

\newcommand{\Magritte}{{\sc Magritte}}

\newcommand{\orcid}[1]{{\hskip.5mm \href{#1}{\includegraphics[height=8px]{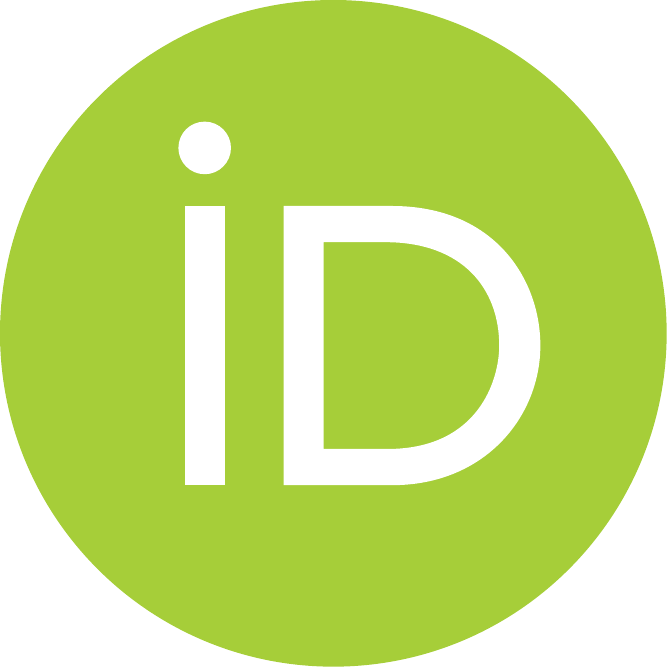}} \hskip.5mm}}

\usepackage{mathtools}
\DeclarePairedDelimiter\ceil{\lceil}{\rceil}

\title
[\textsc{Magritte}: II adaptive ray-tracing and meshing]
{\textsc{Magritte}, a modern software library for 3D radiative transfer: \\
  II. Adaptive ray-tracing, mesh construction and reduction
}

\author
[F. De Ceuster et al.]
{Frederik De Ceuster $^{\orcid{https://orcid.org/0000-0001-5887-8498} 1, 2}$\thanks{Contact e-mail: \href{frederik.deceuster@kuleuven.be}{frederik.deceuster@kuleuven.be}},
 Jan Bolte           $^{\orcid{https://orcid.org/0000-0002-7991-9663} 2   }$,
 Ward Homan          $^{\orcid{https://orcid.org/0000-0001-7314-5081} 2   }$,
 Silke Maes          $^{\orcid{https://orcid.org/0000-0003-4159-9964} 2   }$,
 \newauthor
 Jolien Malfait      $^{\orcid{https://orcid.org/0000-0002-8850-2763} 2   }$,
 Leen Decin          $^{\orcid{https://orcid.org/0000-0002-5342-8612} 2, 3}$,
 Jeremy Yates        $^{\orcid{https://orcid.org/0000-0003-1954-8749} 1   }$,
 Peter Boyle         $^{\orcid{https://orcid.org/0000-0002-8960-1587} 4, 5}$, \newauthor
 and
 James Hetherington  $^{\orcid{https://orcid.org/0000-0001-6993-0319} 6, 7}$
  \\ \\
  $^{1}$ Department of Physics and Astronomy, University College London, Gower Place, London, WC1E 6BT, UK \\
  $^{2}$ Department of Physics and Astronomy, Institute of Astronomy, KU Leuven, Celestijnenlaan 200D, 3001 Leuven, Belgium \\
  $^{3}$ School of Chemistry, University of Leeds, Leeds LS2 9JT, UK \\
  $^{4}$ School of Physics and Astronomy, The University of Edinburgh, Edinburgh EH9 3FD, UK \\
  $^{5}$ Brookhaven National Laboratory, Upton, NY 11973, USA \\
  $^{6}$ Department of Computer Science, University College London, Bloomsburry, London, WC1E 6EA, UK \\
  $^{7}$ The Alan Turing Institute, 96 Euston Road, Kings Cross, London, NW1 2DB, UK
}

\date{Accepted XXX. Received YYY; in original form ZZZ}

\pubyear{2019}

\begin{document}
\label{firstpage}
\pagerange{\pageref{firstpage}--\pageref{lastpage}}
\maketitle

\begin{abstract}
    Radiative transfer is a notoriously difficult and computationally demanding problem. Yet, it is an indispensable ingredient in nearly all astrophysical and cosmological simulations.
    Choosing an appropriate discretization scheme is a crucial part of the simulation, since it not only determines the direct memory cost of the model but also largely determines the computational cost and the achievable accuracy.
    In this paper, we show how an appropriate choice of directional discretization scheme as well as spatial model mesh can help alleviate the computational cost, while largely retaining the accuracy.
    First, we discuss the adaptive ray-tracing scheme implemented in our 3D radiative transfer library \Magritte{}, that adapts the rays to the spatial mesh and uses a hierarchical directional discretization based on \textsc{HEALPix}.
    Second, we demonstrate how the free and open-source software library \textsc{Gmsh} can be used to generate high quality meshes that can be easily tailored for \Magritte{}.
    In particular, we show how the local element size distribution of the mesh can be used to optimise the sampling of both analytically and numerically defined models.
    Furthermore, we show that when using the output of hydrodynamics simulations as input for a radiative transfer simulation, the number of elements in the input model can often be reduced by an order of magnitude, without significant loss of accuracy in the radiation field.
    We demonstrate this for two models based on a hierarchical octree mesh resulting from adaptive mesh refinement (AMR), as well as two models based on smoothed-particle hydrodynamics (SPH) data.
\end{abstract}




\section{Introduction}
\label{sec:introduction}

Radiative transfer plays a critical role in various astrophysical and cosmological processes.
Not only does it determine what we can and cannot observe, it also actively alters the physical and chemical conditions throughout the Universe through radiative pressure, heating and cooling, and via various photo-ionization and photo-dissociation reactions.
Although it is computationally challenging, it is crucial to be able to accurately account for all relevant radiative processes in the wealth of modern astrophysical and cosmological simulations.

The first step in every computer simulation is finding a proper representation for the simulated objects in the model.
Often, this comes down to finding an appropriate discretization for all physical quantities.
This is an essential step, since the number of elements in the discretization will not only determine the direct memory cost of the model but also its computational cost and ultimately the maximal achievable accuracy of the simulation.
Finding an appropriate discretization scheme is thus a question of optimizing the trade-off between accuracy and computational cost.

Well-established radiation transport solvers such as
\textsc{OpenMC}\footnote{See also \href{https://openmc.org/}{openmc.org/}.} \citep{Romano2015},
\textsc{Tripoli-4}\footnote{See also \href{http://www.cea.fr/nucleaire/tripoli-4/}{www.cea.fr/nucleaire/tripoli-4/}.} \citep{Brun2015}, and
\textsc{MCNP}\footnote{See also \href{https://laws.lanl.gov/vhosts/mcnp.lanl.gov/index.shtml}{laws.lanl.gov/vhosts/mcnp.lanl.gov/index.shtml}.} \citep{Werner2018}, which are used for more industrial applications such as nuclear engineering and medical imaging, use surface-based or combinatorial representations for their geometrical models.
Since these are constructed using computer-aided design (CAD) software, they consist of combinations of well-defined shapes (e.g. cuboids, cylinders, spheres), of certain materials, with well-defined boundary surfaces.
As a result, the radiation transport can be considered through one material at a time and rays can be traced from one surface to the next, similar to the rendering techniques used in modern computer graphics \citep[see e.g.][]{Glassner1989}.
The highly accurate representations of the models allow for the highly accurate solutions that are required in these types of applications.

In contrast, in astrophysical and cosmological simulations, one is interested in the radiation field in fluids with continuously varying radiative properties throughout the model.
Therefore, the geometries of these models always have to be discretized before they can be numerically solved, leading to a first inescapable source of numerical error.
The geometries that are used are often inherited from a previous simulation step.
Typically, these radiative transfer solvers are used on top of a hydrodynamics solver to compute for instance the radiative pressure, or they are used to post-process snapshots of hydrodynamics simulations to produce synthetic observations.
In those cases, the radiative transfer solver uses the same geometric model mesh as the hydrodynamics solver, although those meshes are usually only optimised for the latter.

Over the years, the spatial discretization schemes used in hydrodynamics solvers have evolved from static structured meshes, to hierarchical meshes resulting from adaptive mesh refinement \citep[AMR,][]{Berger1989}, to unstructured and dynamically evolving meshes \citep{Springel2010}.
Additionally, there are the mesh-less smoothed-particle hydrodynamics (SPH) solvers that do not rely on a mesh, but rather evolve a set of particles with appropriate smoothing kernels \citep{Lucy1977, Gingold1977}. The spatial discretization schemes used in radiative transfer simulations evolved accordingly from structured to unstructured meshes \citep[see e.g.][]{Ritzerveld2006}, and further to mesh-less schemes \citep[see e.g.][]{Bisbas2012, DeCeuster2019}.
For an assessment of the use of unstructured (Voronoi) meshes in (Monte Carlo) radiative transfer, see e.g. \cite{Camps2013}, and for the use of hierarchical octree and the more general kd-tree meshes see e.g. \cite{Saftly2013, Saftly2014}.

In further contrast to the industrial radiative transfer applications, in astrophysics and cosmology we are almost never interested in a highly accurate solution of a specific model, but e.g. rather in understanding the more general driving mechanisms that govern a set of models.
For instance, where in nuclear engineering it is crucial to be able to accurately describe the effect of adding a single fuel rod in a reactor, it is far less important in cosmology, for instance, to be able to describe the effect of one additional filament in the cosmic web.
This allows us to optimize our models more aggressively, retaining only the essential features under investigation.

This is the second paper in a series on \Magritte{}\footnote{See also \href{https://github.com/Magritte-code}{github.com/Magritte-code}.}: a modern open-source software library for 3D radiative transfer \citep{DeCeuster2019}.
\Magritte{} is a deterministic ray-tracer that uses a formal solver to compute the radiation field along a fixed set of rays (i.e. directions) through the model.
In this paper, we discuss our implementation of an improved ray-tracing scheme, that adapts to the spatial discretization and uses a hierarchical directional discretization based on \textsc{HEALPix}\footnote{\label{HEALPix}See also \href{http://healpix.sourceforge.net}{healpix.sourceforge.net}.} \citep{Gorski2005}.
Furthermore, we demonstrate how the free and open-source software library \textsc{Gmsh}\footnote{See also \href{http://gmsh.info/}{gmsh.info}.} \citep{Geuzaine2009} can be used to efficiently generate high quality meshes that can be used to construct radiative transfer models in \Magritte{}.
Finally, we present and demonstrate a simple algorithm that can be used to reduce the size (i.e. the number of elements) of an input model for radiative transfer simulations by an order of magnitude without a significant loss of accuracy.

The structure of this paper is as follows.
In Section \ref{sec:methods} we discuss the implementation of our adaptive ray-tracing scheme and explain how we use \textsc{Gmsh} to efficiently generate meshes for input models.
Further, in Section \ref{sec:applications}, we demonstrate the mesh generation process with an analytic Archimedean spiral model and four snapshots of hydrodynamics simulations, two of which using AMR and two using SPH.
Finally, in Section \ref{sec:discussion}, we discuss our results and we present our conclusions in Section \ref{sec:conclusion}.

\section{Methods}
\label{sec:methods}

\subsection{Adaptive ray-tracing}
\label{subsec:adaptive_ray_tracing}

Various methods have been devised to compute the radiation field in a given medium by solving the radiative transfer equation. \Magritte{} uses a long-characteristic formal solver and thus solves the radiative transfer equation along a set of predefined rays through the model.
The ray-tracer follows a straight line through the model and gathers the emissivities and opacities along the ray that are then used to solve the radiative transfer equation.
By stepping from one point to the next it indirectly determines the optical depth increments, i.e. the step size, in the discretized transfer equation used by the solver.
The first version of \Magritte{} employed a second-order radiative transfer solver \citep{Feautrier1964, DeCeuster2019}, but the current version also provides the more accurate (Hermitian) fourth-order scheme by \cite{Auer1976}, with second and third-order boundary conditions respectively \citep{Auer1967}.
In the following sections, we describe three ways in which the rays are adapted to further improve the accuracy of the solver.

\subsubsection{Adaptive velocity sampling}
\label{subsubsec:adaptive_velocity_sampling}

Since Doppler shifts can cause significant variations in emissivity and opacity within a frequency bin, it is crucial to carefully sample the velocities encountered along a ray. This becomes even more important for line radiative transfer, where significant changes occur in particularly narrow frequency ranges. In \cite{DeCeuster2019}, it was already discussed how \Magritte{} accounts for this by interpolating the velocity along a ray between two points if its change is too large. In our new version of the solver, we extended this by employing a similar method to the optical depth increments encountered along a ray.

\subsubsection{Adaptive optical depth increments}
\label{subsubsec:adaptive_optical_depth_increments}

In \Magritte{}, the radiative transfer equation is solved in its second-order or Feautrier form \citep{Feautrier1964},
\begin{equation}
    \left(1 \ - \ \frac{\D^{2}}{\D\tau^{2}}\right) u(\x,\n) \ = \ S(\x,\n),
\label{eq:RTE3}
\end{equation}
where $u(\x,\n)$ is the mean intensity along direction $\n$, and $S(\x,\n)$ and $\tau(\x,\n)$ are respectively the effective source function and effective optical depth, which can be derived from the local emissivities and opacities \citep[see][for more details]{DeCeuster2019}.
The optical depth is thus the relevant dependent variable in the differential equation.
Assuming a proper sampling of the emissivity and opacity, the discretization error will thus be determined mainly by the size of the optical depth increments.
This means that a model mesh can perfectly sample the relevant optical data, but nevertheless produce a large discretization error.
To resolve this, we adapt the optical depth increments as they are computed.
In particular, we divide the interval on the ray between the (projected) points $n$ and $n+1$ in $n_{\text{inter}}$ equal parts and linearly interpolate the emissivities and opacities on the sub-intervals. By defining the number of interpolations as
\begin{equation}
   n_{\text{inter}} \ \equiv \ \ceil*{\frac{\max \left\{\chi_{n}, \chi_{n+1}\right\} \Delta s_{n}}{\Delta\tau_{\max}}},
\end{equation}
where $\chi$ denotes the opacity and $\Delta s_{n}$ the distance increment along the ray, we can ensure that the optical depth increments along the ray are always smaller than a predefined value $\Delta\tau_{\max}$.

It is important to note that the adaptive velocity sampling is still a separate process that happens before the adaptive optical depth increments are computed and that the former cannot be included in or replaced by the latter. For example, a large change in velocity along a ray which Doppler shifts a line from one wing to the other will not result in a particularly large optical depth increment, but will leave the line unaccounted for, increasing the error on the computed radiation field.

\subsubsection{Adaptive directional discretization}
\label{subsubsec:adaptive_directional_discretization}

Many applications of radiative transfer computations require directional integrals over the radiation field to compute, for instance, the radiative heating or cooling, or to compute the radiative pressure. In those cases also the directions of the rays need to be discretized.
Given a function $y\left(\x,\n\right)$, the directional integral is discretized as
\begin{equation}
    \oint \D \Omega \ y \left(\x, \n\right) \ \rightarrow \ \sum_{r\in\mathcal{R}_{i}} w_{i, r} \ y_{i, r},
\end{equation}
where the $i$ and $r$ indicate the point and ray index respectively, and $\mathcal{R}_{i}$ is the set of rays originating from point $i$.
The main difficulty is to determine this set of rays ($\mathcal{R}_{i}$) for each point. Once these are known, the corresponding weights $w_{i, r}$ can readily be computed.

Assuming that the points in the mesh properly sample the relevant distributions in the model, for each point in the mesh, the directions of the other points with respect to that point will properly sample the relevant directions for that point.
Therefore, ideally, the discretization of the directions for a point in the mesh would follow the distribution of the directions of the other points with respect to that point.
Furthermore, since this discretization has to be generated for every point in the mesh, the procedure cannot be too computationally demanding.

Therefore, we opted for a structured adaptively refining scheme based on \textsc{HEALPix} \citep{Gorski2005}.
Given a level of refinement, $l$, \textsc{HEALPix} provides a discretization of the unit sphere in $12 \times 4^{l}$ uniformly distributed pixels of equal area.
By stitching together parts of these uniform HEALPix discretizations with different levels of refinement, we can obtain a locally refined directional discretization adapted to the point density in the mesh.

\begin{figure}
  \includegraphics[width=\columnwidth]{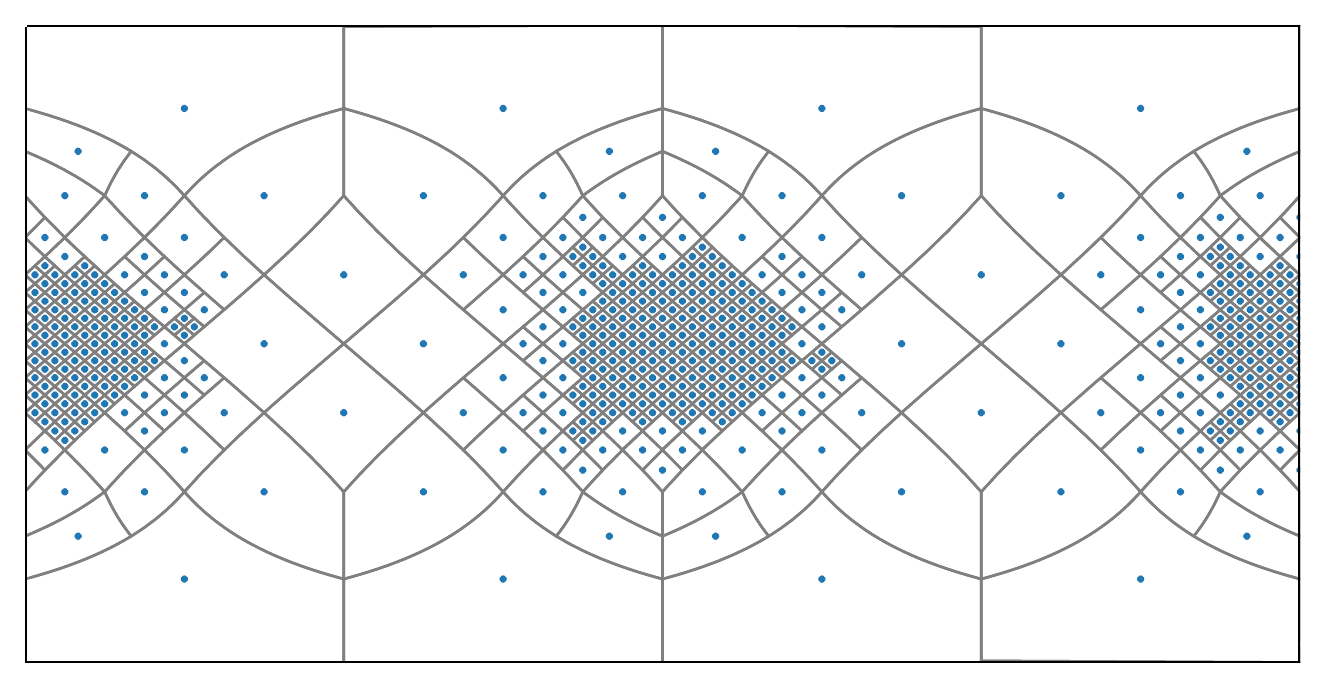}
  \caption{Cartesian projection of an example of the adaptive hierarchical discretization of directions around a point, in this case using four different orders (from $l_{\min}=1$ to $l_{\max}=4$) of the \textsc{HEALPix} scheme. More rays are traced in the directions with a higher mesh point density. The blue dots indicate the centres of the direction vectors and the grey lines delimit the  (HEALPix) pixels.}
  \label{fig:adaptive_ray_tracing}
\end{figure}

We start with a HEALPix discretization with a minimal level of refinement, $l_{\min}$, and refine the pixels according to the distribution of the directions of the other points, until a certain maximal level of refinement, $l_{\max}$, is reached.
Figure \ref{fig:adaptive_ray_tracing} shows a Cartesian projection of an example of the resulting discretization of the unit sphere and the end points of the corresponding direction vectors.
Since \Magritte{} considers pairs of antipodal rays, there is an antipodal symmetry in the directional discretisation.

To obtain the distribution of the directions of the other points with respect to the point under consideration, we draw a uniformly distributed sample of 10 000 points from the mesh and record the \textsc{HEALPix} pixels they belong to for each level of refinement from $l_{\min}$ up to and including $l_{\max}$.
In practice, we only need to compute the distribution at the highest level of refinement.
The results for the other levels can be obtained by downgrading the resulting map, leveraging \textsc{HEALPix}' nested ordering scheme \citep{Gorski2005}.
To decide which pixels to refine, we order them according to the number of directions of other points belonging to them, and only refine the top half.
To ensure that the same number of rays is traced for each point we fix the number of pixels that is refined at each level $l$ to be $6 \times 2^{l}$.
This implies that at each level (except $l_{\max}$), half of the pixels is further refined.

Since \textsc{HEALPix} partitions the unit sphere in pixels of equal area, the corresponding weight for a pixel obtained with a level of refinement $l(i, r)$ is the inverse of the number of pixels at that level
\begin{equation}
    w_{i, r} \ = \ \frac{1}{12 \times 4 ^{l(i, r)}},
\end{equation}
where the level of refinement depends on the originating point $i$ and the direction of the ray $r$. This adaptive refinement scheme allows to sample the directions better with fewer rays.

\subsection{Mesh constructing and reduction}
\label{subsec:meshing}

\begin{figure}
  \includegraphics[width=\columnwidth]{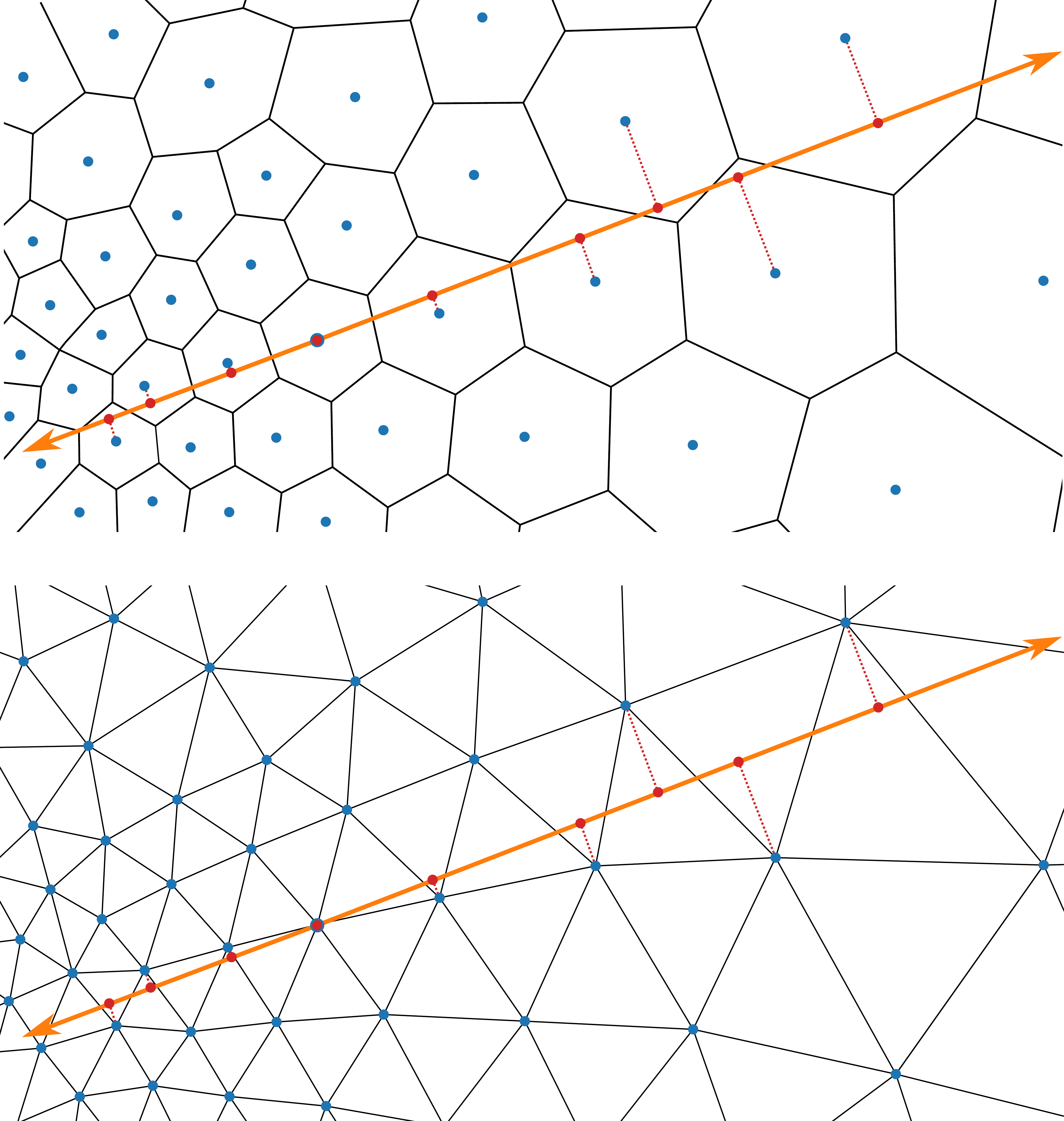}
  \caption{Ray traced through the Voronoi mesh \textit{(top)} and its topologically dual Delaunay mesh \textit{(bottom)}. Note that every distance increment along the ray is the projection of the edge of the Delaunay triangle connecting the traversed Voronoi cell centres. Hence, the edge lengths are a local upper limit for the step size along the ray.}
  \label{fig:ray_tracing}
\end{figure}

Over the years, many different algorithms have been devised to partition a given volume for use in a computer, see for instance the classic treatment by \cite{Thompson1998}, the more recent account by \cite{George2008}, and the references therein.
Many of these algorithms have furthermore been implemented in various software libraries.
For all models in this paper we have used the free and open-source meshing library called \textsc{Gmsh} by \cite{Geuzaine2009}.
\textsc{Gmsh} provides various methods to generate a tetrahedral Delaunay mesh for a domain given a desired local element size distribution.
Since the local element size is directly related to the local edge lengths of the tetrahedra, it allows us to control the step sizes along a ray traced through the domain.
Although \Magritte{} does not require a complete and consistent mesh (it only requires a point cloud and nearest neighbour information), the Delaunay meshes, or their topologically dual Voronoi meshes, provide an excellent means to capture the complex morphologies typically encountered in radiative transfer simulations.
Figure \ref{fig:ray_tracing} shows a ray traced through a domain of Voronoi cells and the corresponding Delaunay tetrahedralization. To use these meshes in \Magritte{}, the Delaunay vertices (or Voronoi centres) can be used as the points and the nearest neighbours can be extracted from the edge list of the mesh, since every pair of nearest neighbours will share an edge.

\subsubsection{Meshing analytic models}
\label{subsubsec:mth_analytic}

Although many astrophysical objects are characterised by irregular structures that are difficult to describe with analytic models, it is nevertheless useful to study analytic models, since they often make it easier to disentangle the effects of the various processes taking place.
When discretizing such a model, the key objective is to properly sample the functions that describe the model parameters.
For simplicity we will restrict ourselves to one function, say $f(\x)$, for which we want to optimise the mesh. We will call this the tracer function.
In radiative transfer computations the density distribution is often used for this purpose.
Properly sampling a function for use in a differential equation solver means properly tracking its changes through the domain.
A mesh that properly samples the tracer function will have small elements whenever the change in the tracer is large and vice versa.
Therefore, in order to quantify the desired local element size distribution, we need to quantify the maximal local relative change in the tracer function.
This is given by the norm of the gradient of its logarithm, which we will denote as,
\begin{equation}
G f (\x) \ \equiv \ \max_{\n \in S^{2}} \big\{ \n \cdot \nabla f(\x) \ / \ f(\x) \big\} \ = \ \| \nabla \log f(\x) \|.
\end{equation}
This can be used in a map to obtain the desired element size distribution $\ell(\x)$ of the model mesh.
We thus look for a continuous mapping that (at least roughly) maps $\max \{G f(\x)\}$ to $\min \{ \ell (\x)\}$ and $\min \{G f(\x)\}$ to $\max \{ \ell (\x)\}$.

Given the form of the radiative transfer equation one could argue that in constructing the mesh, one should strive to keep the optical depth increments as small as possible.
However, this is not strictly required here, since \Magritte{} will automatically limit the size of the optical depth increments by interpolating the optical properties where necessary. Furthermore, it was already shown, in the context of subdivision stopping criteria for adaptively refined meshes by \cite{Saftly2014}, that it is far more important that a mesh allows to accurately sample the model functions than to limit the encountered optical depth increments.
Therefore, we use a linear mapping from $\left[\max \{G f(\x)\}, \min \{G f(\x)\}\right]$ to $\left[\min \{ \ell (\x)\}, \max \{ \ell (\x)\}\right]$.
Any other mapping would have a larger local gradient in the desired local element size function and would therefore make it harder to mesh.

Although \textsc{Gmsh} has the option to construct meshes from an analytically defined element size distribution, it is often much simpler to provide the element size distribution evaluated on a background mesh. Therefore, we consider a regular Cartesian mesh that will be used as a background mesh. The resolution of the background mesh is determined by the smallest scales of the tracer function that we want to resolve, say $\ell_{\min}$. This is also the lower bound for the desired element size distribution. Similarly, we define an upper bound for the desired element size distribution, $\ell_{\max}$, which can often be conveniently defined as a fraction of the size of the domain.

Once an appropriate background mesh is created with the desired element size distribution evaluated on it, \textsc{Gmsh} can accordingly generate a mesh for a given domain.

\subsubsection{Re-meshing existing models}
\label{subsubsec:mth_remesh}

Since radiative transfer simulations are often only a component in a bigger simulation pipeline, the spatial discretizations that are used are often inherited from previous simulation steps.
The corresponding meshes are usually not tailored to the radiative transfer solvers and as a result contain an exceedingly large amount of elements.
Here we present a simple algorithm to reduce the number elements in a given mesh, while preserving a proper sampling of a given tracer function.

When re-meshing a given model, we need to know where the resolution of the original mesh is essential for the accurate representation of the model and where it could be coarsened. This can be quantified by the maximum relative change of the tracer function at a point with respect to its neighbours. This can be expressed as an operator acting on the tracer function
\begin{equation}
    Gf_{i} \ \equiv \ \max \left \{ \left| \frac{f_{i} - f_{n}}{f_{i} + f_{n}} \right|, \ \text{for each neighbour} \ n \ \text{of point} \ i \right \}.
\end{equation}
By definition, $Gf_{i} \in [0, 1]$, so we can define a simple threshold value, $G_{\text{thres}}$, above which the original local element size is deemed essential.
We can now assign a desired local element size $\ell(\x)$ as a fraction of the original local element sizes $L(\x)$, where the fraction is determined by the local change in the tracer function
\begin{equation}
    \ell(\x) \ = \ L(\x) \
    \begin{cases}
        f_{\text{small}} & \text{if} \ G f(\x) \geq G_{\text{thres}} \\
        f_{\text{large}} & \text{otherwise}
    \end{cases}
\label{eq:map}
\end{equation}
where the local element sizes of the original mesh are given by
\begin{align}
    L_{i} \ \equiv \ \text{mean}\big\{ \left\|\x _{i} - \x _{n} \right\|, \ \text{for each neighbour} \ n \ \text{of point} \ i \big\} .
\end{align}
The algorithm thus depends on three parameters ($f_{\text{small}}$, $f_{\text{large}}$, and $G_{\text{thres}}$), for which the values used for the applications in this paper can be found in Table \ref{tab:alg_params}.

Several variations are possible on the mapping to obtain the desired local element size distribution.
However, due to the stochastic nature of the mesh construction process, the differences quickly blur resulting in similar meshes.
The particular map presented here (\ref{eq:map}) was chosen because it is the direct mathematical representation of our objective to coarsen the mesh where possible and retain the original mesh size where it is deemed essential for a proper representation of the model.

\begin{table}
	\centering
	\caption{Empirically determined parameters for the reduction algorithm.}
	\label{tab:alg_params}
	\begin{tabular}{l r r r}
	    \hline
        Model type & $f_{\text{small}}$ & $f_{\text{large}}$ & $G_{\text{thres}}$ \\ \hline
        AMR        &               0.90 &               2.10 &               0.10 \\
        SPH        &               1.00 &               2.15 &               0.21 \\
	\end{tabular}
\end{table}

\begin{table}
	\centering
	\caption{Properties of the original and reduced meshes and the resulting speedup that is achieved in computing the radiation field.}
	\label{tab:meshes}
	\begin{tabular}{l r r r r}
	    \hline
        Model       & $N_{\text{original}}$ & $N_{\text{reduced}}$ & reduction & speedup \\ \hline
        AMR regular &               642 048 &               62 984 &      10.2 &    18.3 \\
        AMR erratic &               627 712 &               75 137 &       8.4 &    12.6 \\
        SPH regular &               916 601 &               82 554 &      11.1 &    35.0 \\
        SPH erratic &               820 471 &               76 660 &      10.7 &    34.1 \\
	\end{tabular}
\end{table}

\section{Applications}
\label{sec:applications}

In this section we apply the mesh construction and reduction methods described above to a set of models inspired by analytic and numeric models of spiral-shaped stellar outflows.
These types of models are currently being developed to investigate the effect of companions on the shapes of the dust-driven winds of cool evolved stars, which have been found to deviate substantially from the originally assumed spherically symmetric wind model \cite{Decin2020}.
Understanding the origin of these features could help explain the morphological evolution towards the highly aspherical planetary nebula phase.

\subsection{Meshing analytic models}
\label{subsec:app_analytic}

\begin{figure}
	\centering
	\includegraphics[width=\columnwidth]{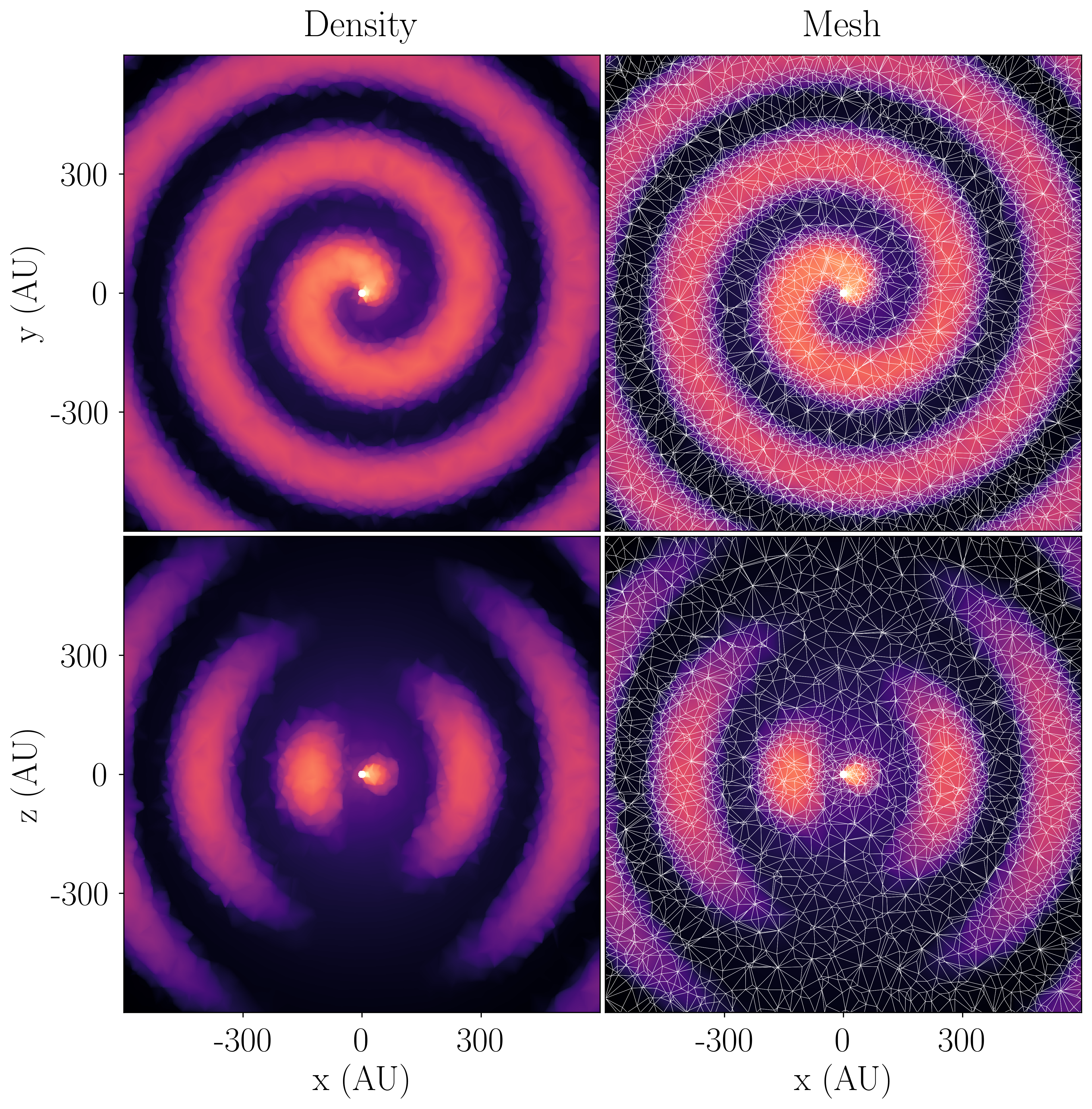}
  \caption{Model mesh for the analytic stellar wind model described by an Archimedean spiral. The top row shows slices through the centre along the $xy$-plane and the bottom row shows slices along the $xz$-plane.}
  \label{fig:analytic_spiral}
\end{figure}

As an example of an analytically defined model, we consider a stellar wind described by an Archimedean spiral with generic parameters, following \cite{Homan2015}.
The full details of the model together with a notebook implementation can be found online\footnote{See \href{https://github.com/Magritte-code/Examples}{github.com/Magritte-code/Examples}.}.
Figure \ref{fig:analytic_spiral} shows two slices through the model, showing the density distribution as well as the underlying mesh generated with \textsc{Gmsh} using the method described in Section \ref{subsubsec:mth_analytic}.
In this example, the density was used as a tracer function to determine the desired local element size distribution, with a minimum desired element size of $\ell_{\min} = 15 \text{\ AU}$ and a maximum desired element size of $\ell_{\max} = 50 \text{\ AU}$.
The regular Cartesian background mesh is defined in a cubic box of size $(1200 \text{\ AU})^{3}$ and a resolution of $100^{3}$ elements.
The resulting mesh consists of 49 347 points, a modest number considering the relatively complex morphology.
One can easily obtain even sparser meshes, either by increasing the minimum or maximum desired element sizes, or by applying the method presented in Section \ref{subsubsec:mth_remesh}. The latter technique will be demonstrated in the next section.

\subsection{Re-meshing existing models}
\label{subsec:app_remesh}

In the following we will consider four specific examples of how hydrodynamics models can be reduced or coarsened before they are used as input for a radiative transfer solver. We consider two models based on an hierarchical octree mesh resulting from adaptive mesh refinement (AMR), and two models based on smoothed-particle hydrodynamics (SPH) simulation data.

\subsubsection{AMR models}

The idea of an octree discretization of a 3D space is to locally subdivide an initial cubic cell into eight sub-cells until the desired local mesh size is achieved.
Hydrodynamics solvers using adaptive mesh refinement (AMR) often use an octree as underlying geometric data structure.
We consider late snapshots of two hydrodynamics models of the intricate stellar outflow produced by a mass-losing asymptotic giant branch (AGB) star as it is perturbed by a companion, modelled using \textsc{MPI-AMRVAC}\footnote{See also \href{https://amrvac.org/}{amrvac.org}.} \citep{Xia2018}.
We used the code with a Cartesian mesh and allowed for 8 levels of adaptive refinement.

\begin{figure*}
	\centering
	\includegraphics[width=2.1\columnwidth]{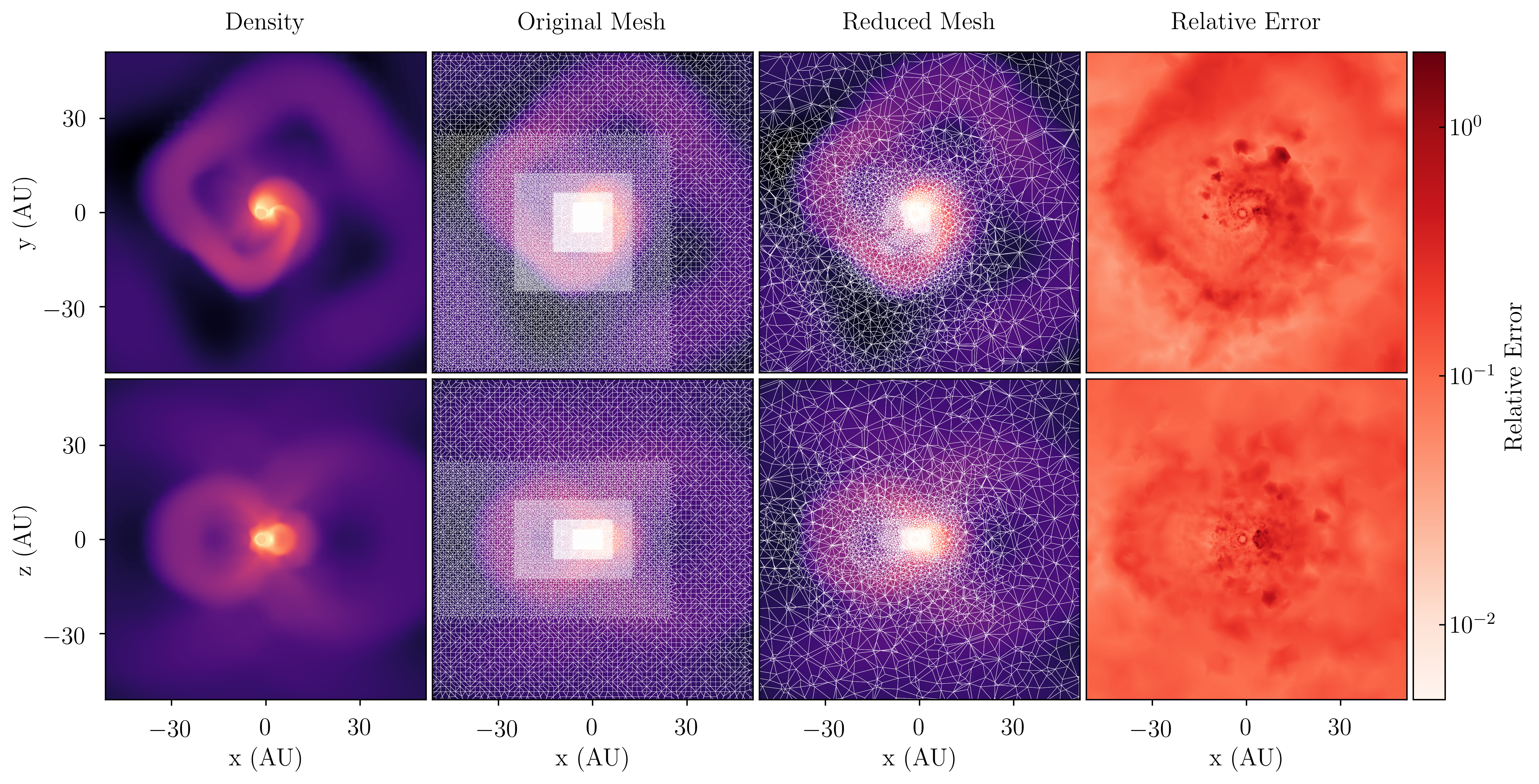}
  \caption{Comparison between the original and reduced model mesh for the octree version of the regular spiral model. The properties of the meshes can be found in Table \ref{tab:meshes}. The top row shows slices through the centre along the $xy$-plane and the bottom row shows slices along the $xz$-plane. The relative error in the rightmost column is computed as the average over all directions and frequency bins of the absolute relative difference between the radiation field computed on the original mesh and the radiation field computed on the reduced mesh when interpolated to the original.}
  \label{fig:amr_regular_meshes_and_errors}
\end{figure*}

\begin{figure*}
	\centering
	\includegraphics[width=2.1\columnwidth]{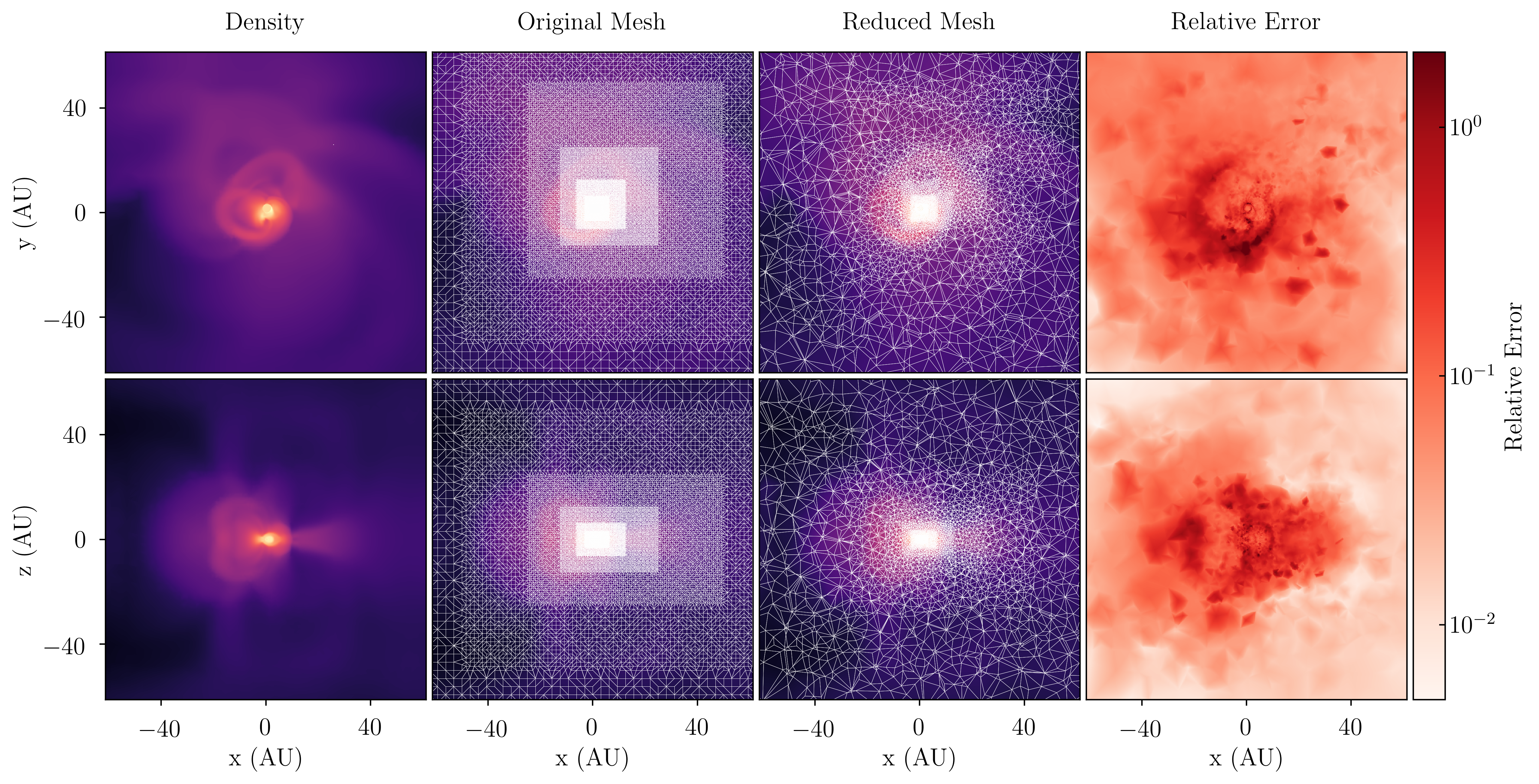}
  \caption{Comparison between the original and reduced model mesh for the octree version of the erratic spiral model. The properties of the meshes can be found in Table \ref{tab:meshes}. The top row shows slices through the centre along the $xy$-plane and the bottom row shows slices along the $xz$-plane. The relative error in the rightmost column is computed as the average over all directions and frequency bins of the absolute relative difference between the radiation field computed on the original mesh and the radiation field computed on the reduced mesh when interpolated to the original.}
  \label{fig:amr_erratic_meshes_and_errors}
\end{figure*}

The first example, shown in Figure \ref{fig:amr_regular_meshes_and_errors}, contains a relatively regular spiral outflow.
Reducing the model using \textsc{Gmsh} and the algorithm described in Section \ref{subsubsec:mth_remesh}, the resulting reduced mesh contains about 10.2 times fewer points than the original one.
This results in a speedup of 18.3 for the computation of the radiation field.
In the second example, shown in Figure \ref{fig:amr_erratic_meshes_and_errors}, we consider a more erratic spiral outflow.
This leads to a reduced mesh containing 8.4 times fewer points than the original one, which results in a 12.6x speedup.
The parameters of the reduction algorithm can be found in Table \ref{tab:alg_params} and properties of the original and reduced meshes are summarised in Table \ref{tab:meshes}.

In both examples, one can see in the reduced meshes some artefacts of the levels of refinement in the original meshes.
This is due to the fact that the desired element sizes in the algorithm are determined by the original element sizes.

\begin{figure}
	\centering
	\includegraphics[width=\columnwidth]{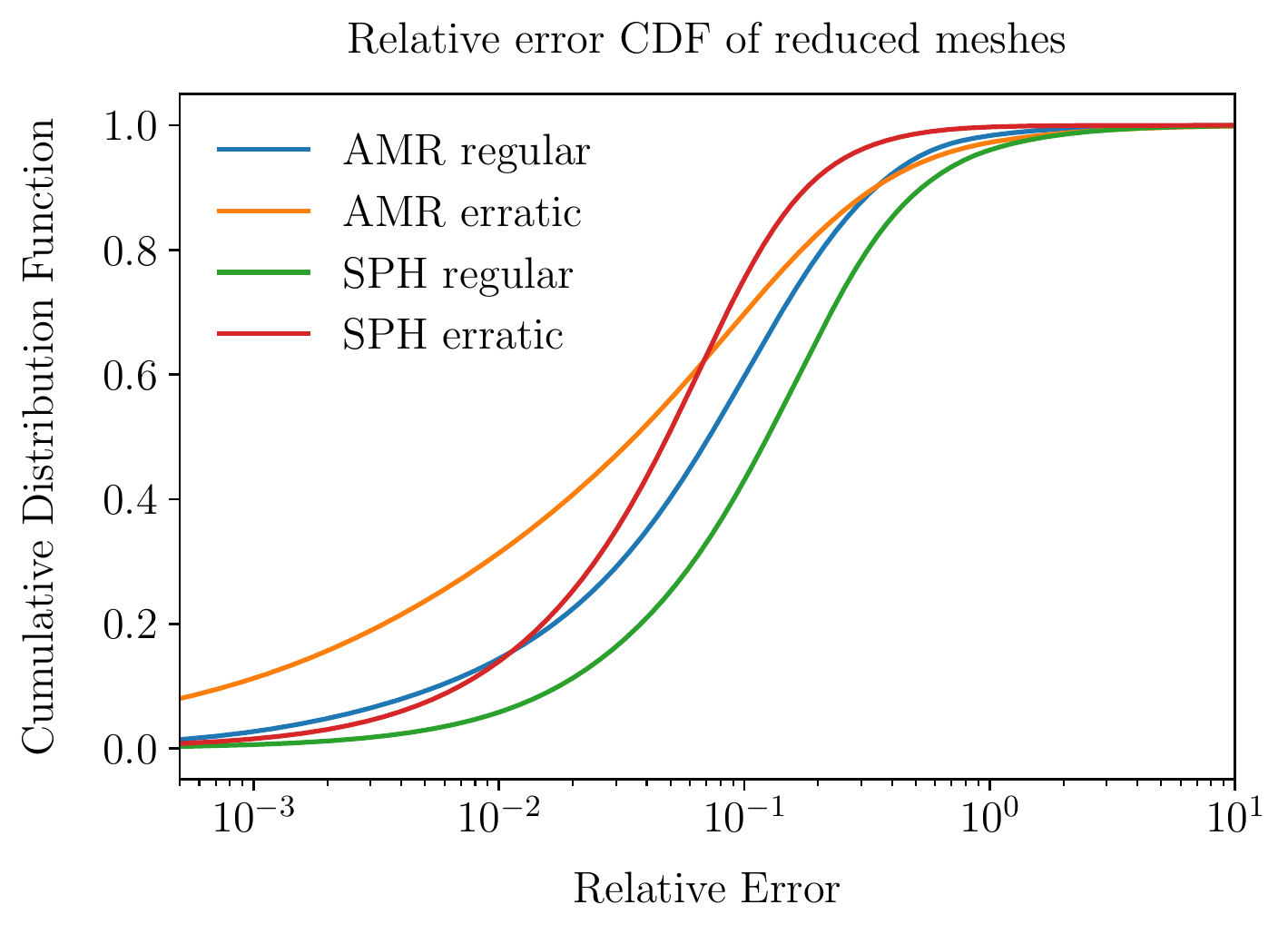}
  \caption{Cumulative distribution function of the relative errors with respect to the original meshes for the different models, computed for 1000 bins. The properties of the different meshes can be found in Table \ref{tab:meshes}.}
  \label{fig:error_hist}
\end{figure}

To quantify the quality of the reduced meshes we compute the radiation field for both the original and reduced models using \Magritte{} and calculate the absolute relative difference between the results.
The solution on the reduced mesh can be mapped to the original by barycentric interpolation on the reduced mesh to each point in the original mesh.
This can readily be done using the \texttt{LinearNDInteprolator} in \textsc{scipy} \citep{Virtanen2019}.
The absolute relative difference between the results can then be computed by point-wise dividing the absolute difference by the result on the original mesh.
In order to gauge the overall distribution of the errors, Figure \ref{fig:error_hist} presents the cumulative density distribution of the relative errors.
The more than 10\% of points with an error below $10^{-3}$ in the AMR erratic model are due to the fact that the reduced mesh at small resolutions still closely resembles the original one.

Since \Magritte{}'s internal geometric data structure consists of a point cloud with nearest neighbour information, the hierarchical octree mesh produced by \textsc{MPI-AMRVAC} cannot be used directly as input for \Magritte{}.
However, a natural way to map the octree mesh to a point cloud is to associate all cell data with the cell centre, use the cell centres as points and extract the nearest neighbour information from the hierarchical octree.

\begin{figure*}
	\centering
	\includegraphics[width=2.1\columnwidth]{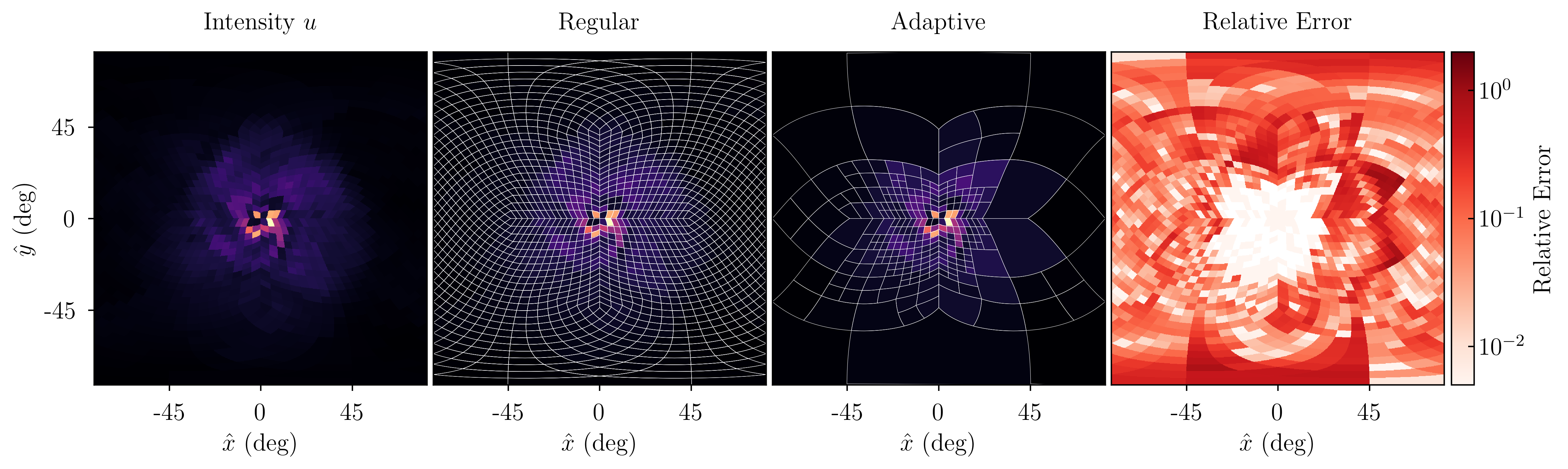}
  \caption{Comparison between a regular and an adaptive discretization of the directions for a point located at $(x,y,z) = (0,0,42  )$ AU in the regular spiral AMR model. The point and rotation are chosen such that the viewing angle resembles the slice in the top row of Figure \ref{fig:amr_regular_meshes_and_errors}. Since the mean intensity along a ray ($u$) is symmetric, each plot shows only half of a Cartesian projection.}
  \label{fig:amr_regular_rays_and_errors}
\end{figure*}

\begin{figure*}
	\centering
	\includegraphics[width=2.1\columnwidth]{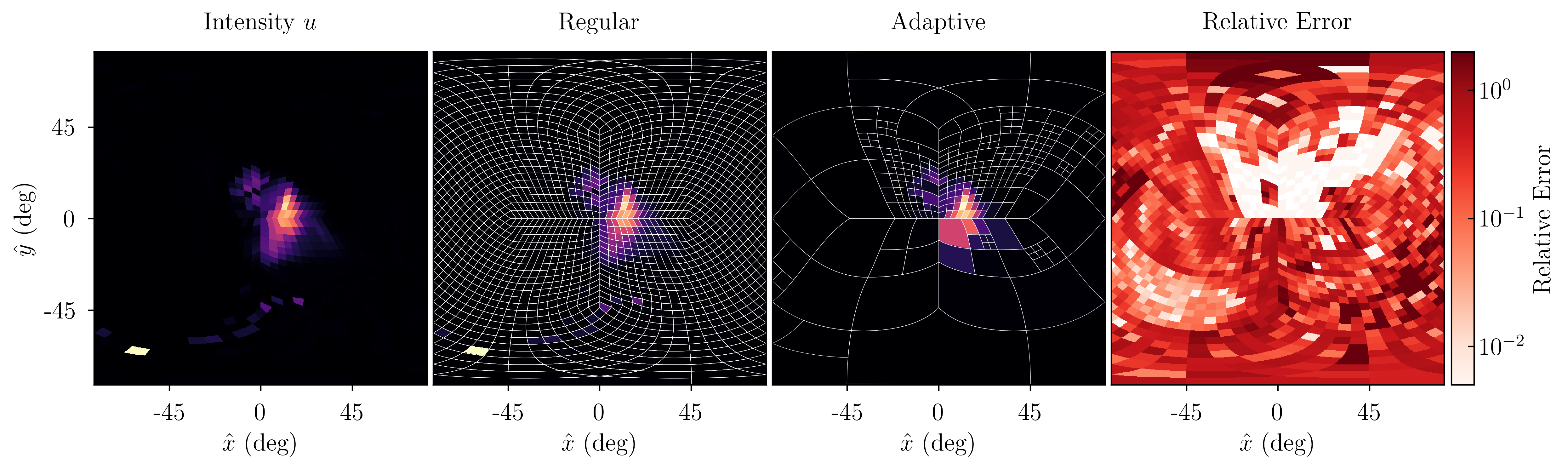}
  \caption{Comparison between a regular and an adaptive discretization of the directions for a point located at $(x,y,z) = (0,0,6)$ AU in the erratic spiral AMR model. The point and rotation are chosen such that the viewing angle resembles the slice in the top row of Figure \ref{fig:amr_erratic_meshes_and_errors}. Since the mean intensity along a ray ($u$) is symmetric, each plot shows only half of a Cartesian projection.}
  \label{fig:amr_erratic_rays_and_errors}
\end{figure*}

\begin{figure}
	\centering
	\includegraphics[width=\columnwidth]{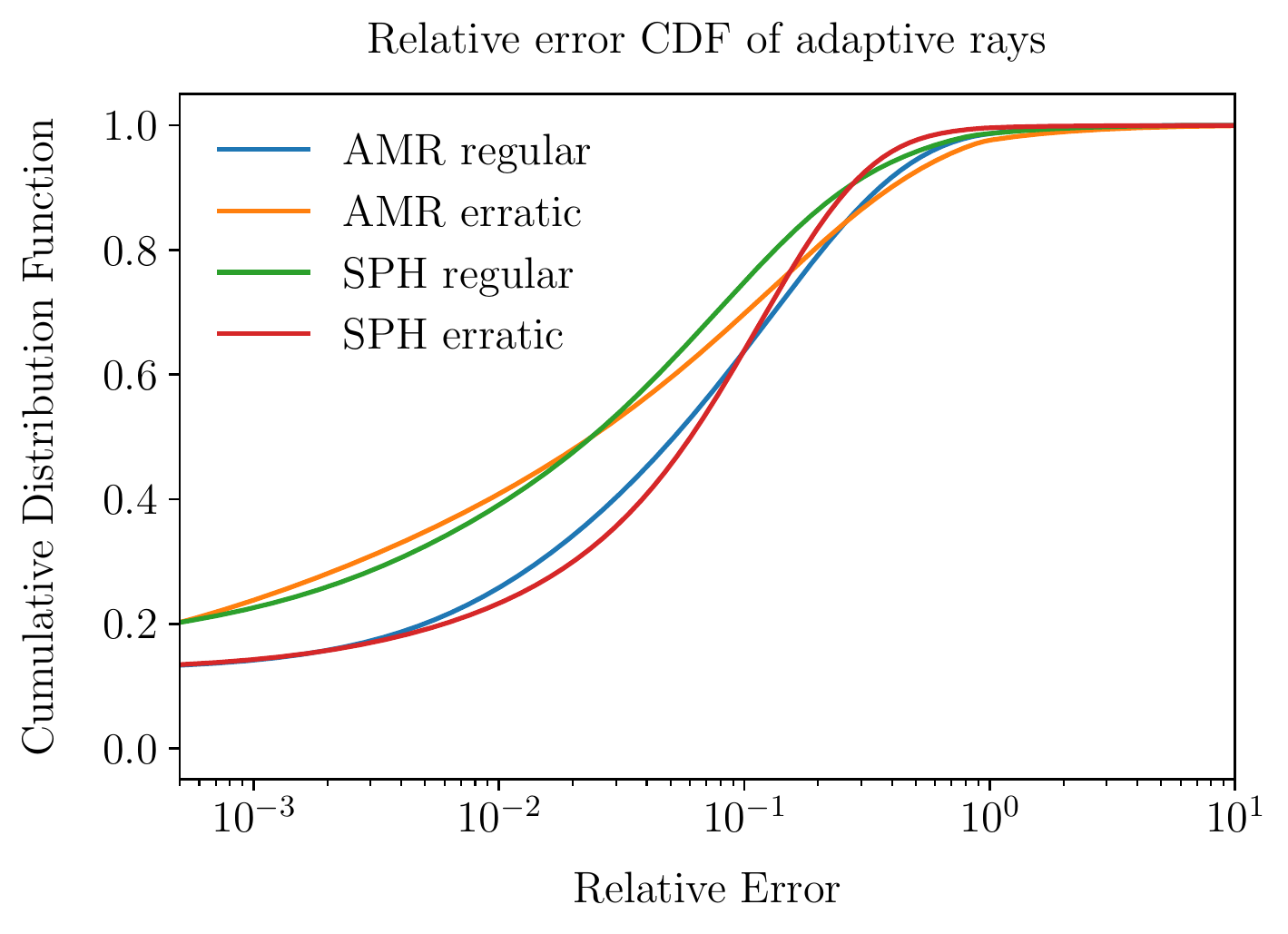}
  \caption{Cumulative distribution function of the relative errors of the adaptively ray-traced models with respect to regular ones, computed for 1000 bins. The reduced meshes (see Table \ref{tab:meshes}), were used as spatial discretization.}
  \label{fig:error_hist_rays}
\end{figure}

Figures \ref{fig:amr_regular_rays_and_errors} and \ref{fig:amr_erratic_rays_and_errors} show a comparison between a dense regular directional discretization containing $12\times2^{8} = 3072$ rays and our adaptive scheme containing 552 rays for a point half way along the $z$-axis looking down on the $xy$-plane of the AMR models.
The adaptive scheme allows for 3 levels of refinement from $l_{\min} = 1$ up until $l_{\max}=4$.
Although the result on the coarser adaptive discretization clearly shows some differences with the finer regular one, the overall relative errors are limited, as can be seen from the cumulative distribution of the relative errors between the dense regular and adaptive models shown in Figure \ref{fig:error_hist_rays}.
More than 60\% of all the rays originating from all the points show a relative error below 10\%, and more than 80\% show a relative error below 20\%.
Note that about 20\% of the points have a relative error below $10^{-3}$.
This is due to the fact that the maximally refined directions in the adaptive scheme have the same order and hence exactly the same rays as in the dense regular scheme, yielding exactly the same results and a negligible error.

\subsubsection{SPH models}

In smoothed-particle hydrodynamics (SPH) simulations, rather than defining the physical quantities on a mesh, the model is described by a number of particles with definite properties and smoothing kernels describing their proliferation.
We again consider late snapshots of two hydrodynamics models of the intricate stellar outflow produced by a mass-losing asymptotic giant branch (AGB) star as it is perturbed by a companion, this time modelled using the smoothed-particle hydrodynamics code \textsc{Phantom}\footnote{See also \href{https://phantomsph.bitbucket.io/}{phantomsph.bitbucket.io}.} \citep{Price2018}.

\begin{figure*}
	\centering
	\includegraphics[width=2.1\columnwidth]{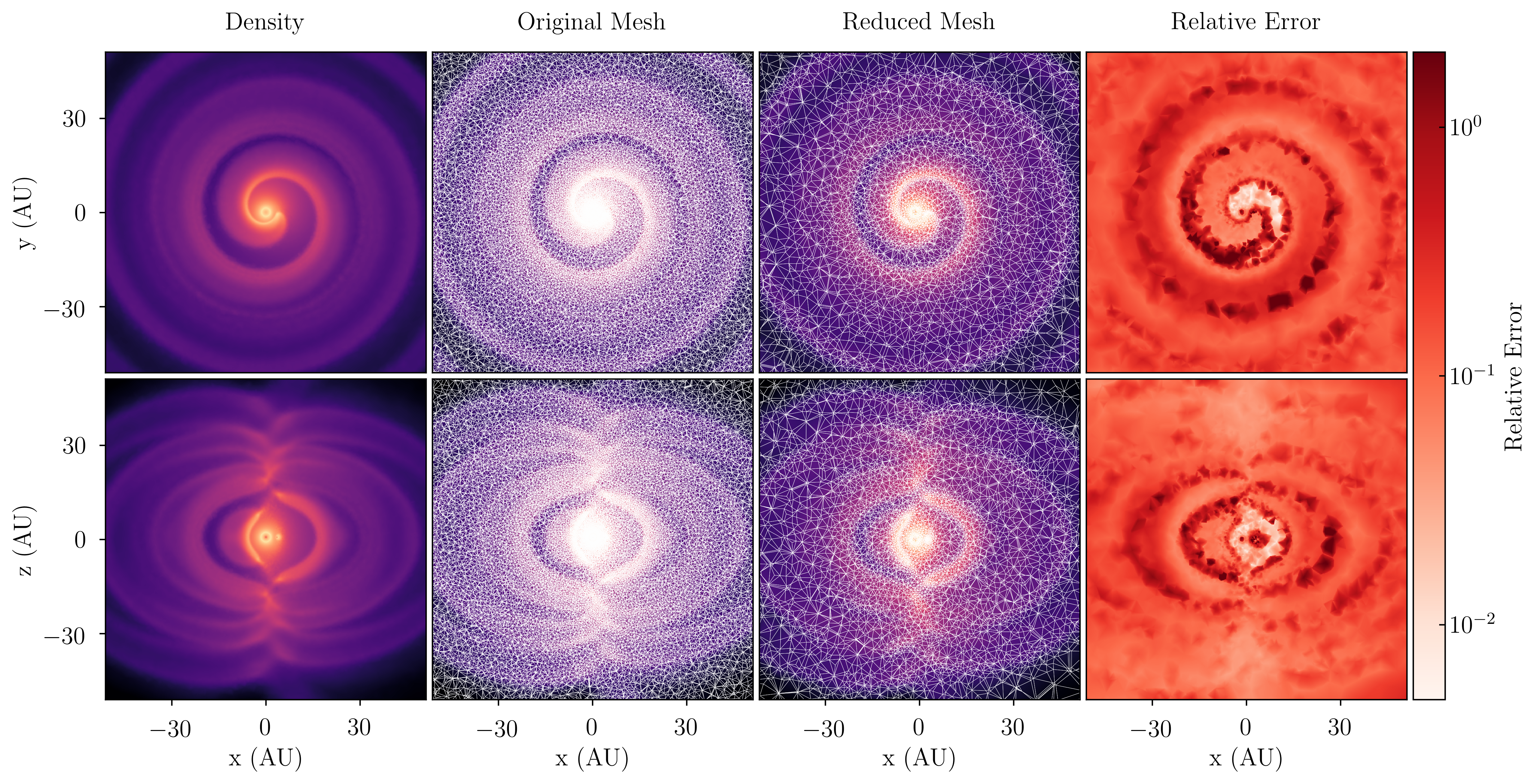}
  \caption{Comparison between the original and reduced model mesh for the SPH version of the regular spiral model. The properties of the meshes can be found in Table \ref{tab:meshes}. The top row shows slices through the centre along the $xy$-plane and the bottom row shows slices along the $xz$-plane. The relative error in the rightmost column is computed as the average over all directions and frequency bins of the absolute relative difference between the radiation field computed on the original mesh and the radiation field computed on the reduced mesh when interpolated to the original.}
  \label{fig:sph_regular_meshes_and_errors}
\end{figure*}

\begin{figure*}
	\centering
	\includegraphics[width=2.1\columnwidth]{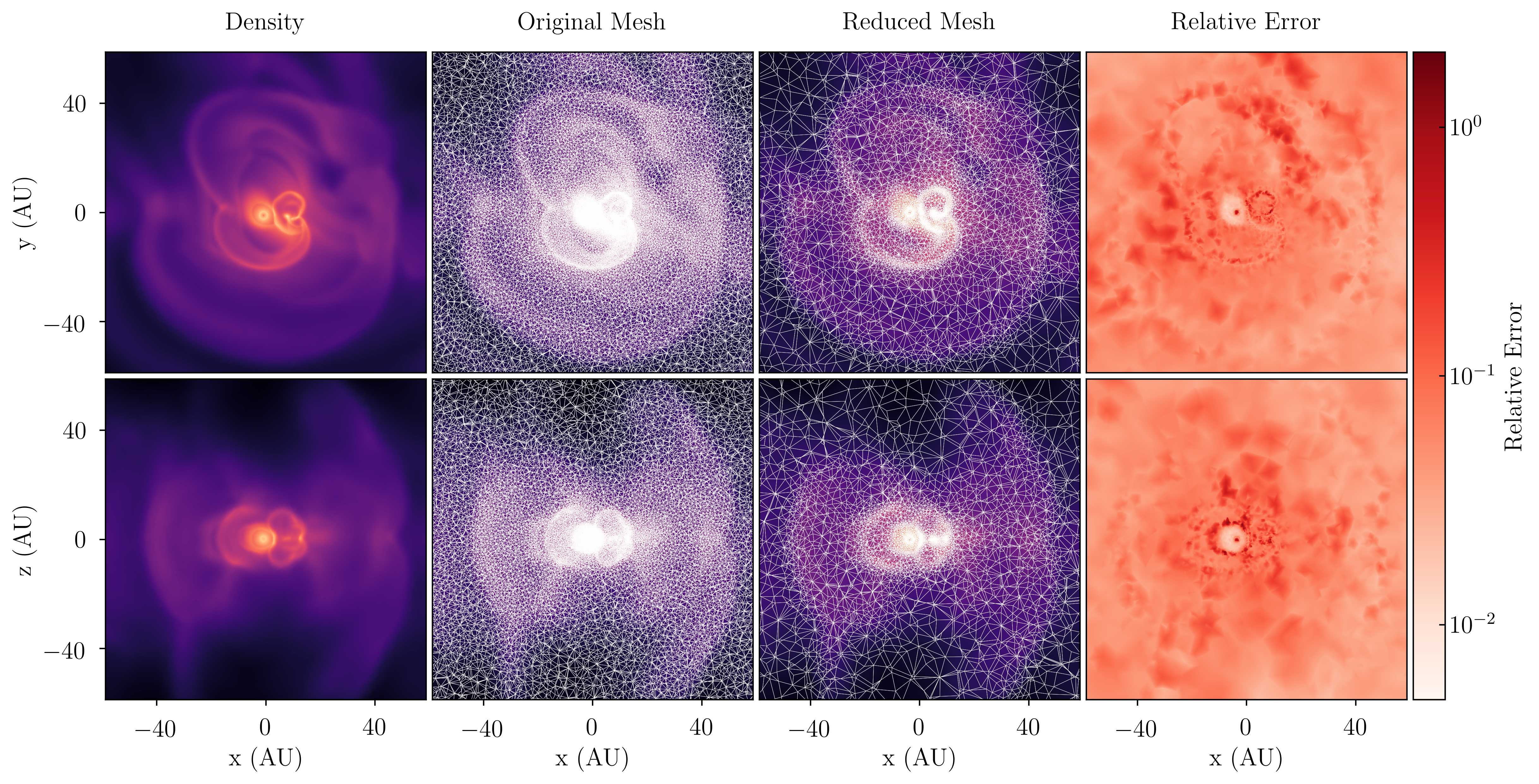}
  \caption{Comparison between the original and reduced model mesh for the SPH version of the erratic spiral model. The properties of the meshes can be found in Table \ref{tab:meshes}. The top row shows slices through the centre along the $xy$-plane and the bottom row shows slices along the $xz$-plane. The relative error in the rightmost column is computed as the average over all directions and frequency bins of the absolute relative difference between the radiation field computed on the original mesh and the radiation field computed on the reduced mesh when interpolated to the original.}
  \label{fig:sph_erratic_meshes_and_errors}
\end{figure*}

The first example, shown in Figure \ref{fig:sph_regular_meshes_and_errors}, describes a very regular spiral outflow.
Reducing the model using \textsc{Gmsh} and the algorithm described in Section \ref{subsubsec:mth_remesh}, the resulting reduced mesh contains 11.1 times fewer points than the original.
This results in a speedup of 35.0 for the computation of the radiation field.
In the second example, shown in Figure \ref{fig:sph_erratic_meshes_and_errors}, we consider a much more erratic spiral outflow.
Despite the complex morphology, the reduced mesh still contains 10.7 times fewer points than the original, resulting in a 34.1x speedup.
The parameters of the reduction algorithm can be found in Table \ref{tab:alg_params} and the properties of the original and reduced meshes are summarised in Table \ref{tab:meshes}.

The quality of the meshes can again be quantified by comparing the results of a radiative transfer computation using \Magritte{} between the original and reduced meshes. The results of the reduced mesh can be mapped to the original one in the same way as with the AMR models.
Figure \ref{fig:error_hist} shows the cumulative distribution of the relative errors between the original and reduced meshes.
The SPH erratic model shows a relative error below 10\% for about 90\% of its points.
This can be attributed the fact that the original SPH model already had a higher sampling that follows the morphology more closely.
Since the desired element sizes are based on this sampling, this will result in higher quality meshes.

The point cloud structure of an SPH data set maps naturally to \Magritte{}'s internal geometric data structure.
However, it should be noted that in this way we do not account for the smoothing kernels.

\begin{figure*}
	\centering
	\includegraphics[width=2.1\columnwidth]{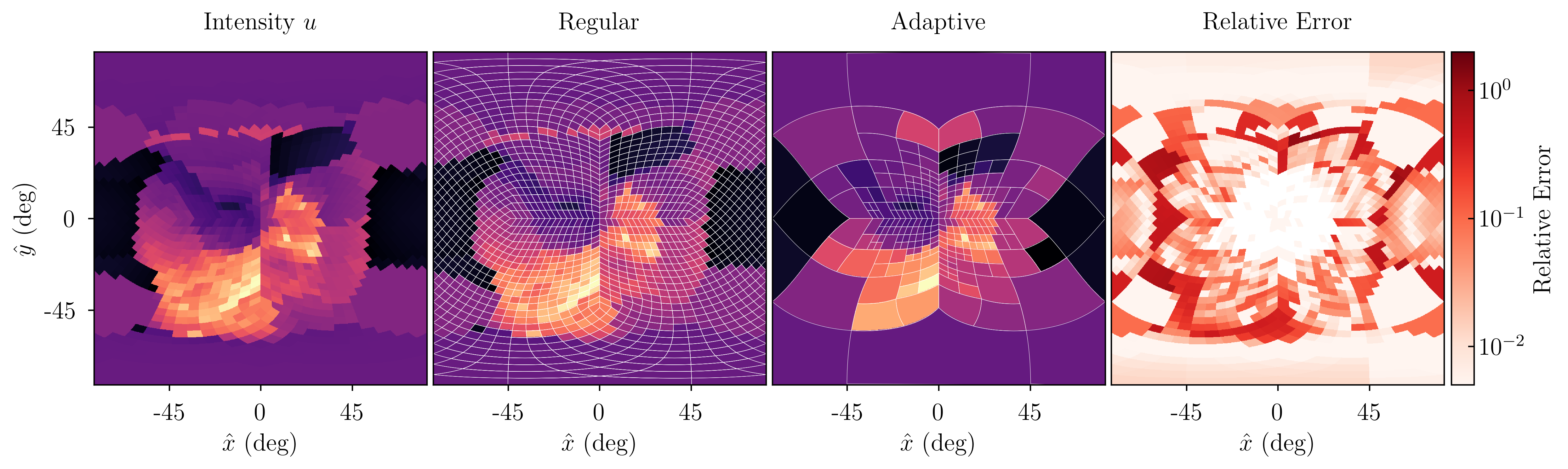}
  \caption{Comparison between a regular and an adaptive discretization of the directions for a point located at $(x,y,z) = (0,0,81)$ AU in the regular spiral SPH model. The point and rotation are chosen such that the viewing angle resembles the slice in the top row of Figure \ref{fig:sph_regular_meshes_and_errors}. Since the mean intensity along a ray ($u$) is symmetric, each plot shows only half of a Cartesian projection.}
  \label{fig:sph_regular_rays_and_errors}
\end{figure*}

\begin{figure*}
	\centering
	\includegraphics[width=2.1\columnwidth]{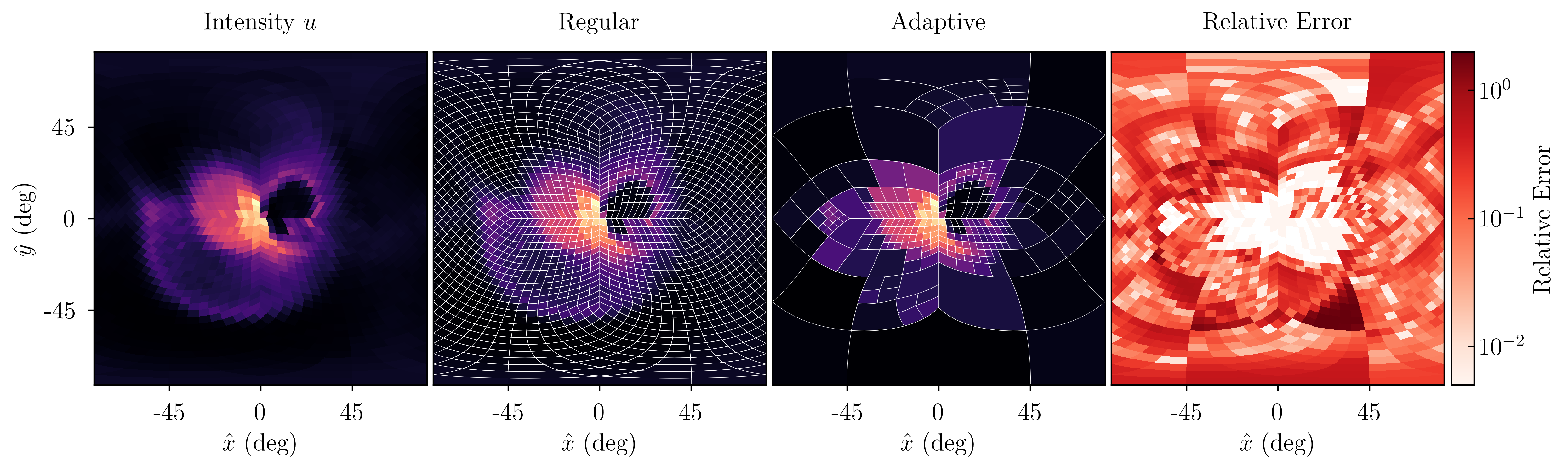}
  \caption{Comparison between a regular and an adaptive discretization of the directions for a point located at $(x,y,z) = (0,0,18)$ AU in the erratic spiral SPH model. The point and rotation are chosen such that the viewing angle resembles the slice in the top row of Figure \ref{fig:sph_erratic_meshes_and_errors}. Since the mean intensity along a ray ($u$) is symmetric, each plot shows only half of a Cartesian projection.}
  \label{fig:sph_erratic_rays_and_errors}
\end{figure*}

Figures \ref{fig:sph_regular_rays_and_errors} and \ref{fig:sph_erratic_rays_and_errors} show a comparison between a dense regular directional discretization containing $12\times2^{8} = 3072$ rays and our adaptive scheme containing 552 rays for a point half way along the $z$-axis looking down on the $xy$-plane of the SPH models.
The adaptive scheme allows for 3 levels of refinement from $l_{\min} = 1$ up until $l_{\max}=4$.
In accordance with the results for the AMR models, the result for the SPH models on the coarser adaptive discretization clearly shows some differences with the finer regular one, while the overall relative errors are again limited, as can be seen from the cumulative distribution of the relative errors between the dense regular and adaptive models shown in Figure \ref{fig:error_hist_rays}.
More than 60\% of all the rays originating from all the points show a relative error below 10\%, and more than 80\% show a relative error below 20\%.
Note also here that about 20\% of the points have a relative error below $10^{-3}$, which can be attributed to the fact that the maximally refined directions in the adaptive scheme have the same order and hence exactly the same rays as in the dense regular scheme, yielding exactly the same results and a negligible error.

\section{Discussion}
\label{sec:discussion}

\subsection{Adaptive ray-tracing}

We should point out that our approach is quite different from \cite{Abel2002}, who originally coined the term ``adaptive ray-tracing'' in the context of radiative transfer in cosmological simulations. Their idea, to split rays as they reach further away from their origin to obtain a more constant volume coverage, is ideal for direct solvers or short-characteristics methods but would be difficult to implement in our second-order solver. Our approach is more related to adaptive mesh refinement (AMR) methods, but applied on-the-fly along the ray and in the directional discretization.

\subsection{Interpolating between meshes}

The choice of interpolation method between coarser and finer meshes is crucial in the determination of the error caused by the mesh reduction. In this paper we used a barycentric scheme implemented in  \textsc{scipy}'s \texttt{LinearNDInteprolator} to interpolate between coarser and finer meshes \citep{Virtanen2019}. However, using for instance a more primitive mapping to nearest neighbours yielded significantly worse results, increasing the relative errors by about a factor 4. Higher order schemes might provide better results but would infer a higher computational cost. Since the barycentric scheme is already quite expensive (it requires roughly about 10\% of the time to generate a mesh reduction) it appears to be the most suitable approach for now.

\subsection{Model symmetries, 1D and 2D ray-tracers}

When a model possesses a symmetry, it is most advantageous to use a solver that can leverage that symmetry.
For instance, a spherically symmetric model can be solved most efficiently using a solver that is effectively 1D.
Similarly, models that posses a cylindrical symmetry, such as disks and bi-polar outflows, can be solved most efficiently using a solver that is effectively 2D. Nevertheless, we tried to generate 3D meshes for models possessing a symmetry (spherical or cylindrical), to see if we could use our methods to generate relatively sparse 3D meshes for these effectively lower dimensional models.
However, since \Magritte{} internally uses a Cartesian co-ordinate system, it turned out that each of these models required an objectionable amount of points to properly represent these models in 3D.
To accommodate that, we equipped \Magritte{} with a dedicated 1D (spherical symmetric) and 2D (cylindrical symmetric) ray-tracer.
The methods outlined in this paper can readily be applied to generate lower dimensional (1D or 2D) models that can now also be processed with \Magritte{}.

\subsection{Future work}



Now we have a set of methods at our disposal that allow us to easily generate meshes tailored to radiative transfer calculations, we can leverage these techniques, for instance, to develop multi-physics multi-grid methods for the iterative solvers in \Magritte{} that aim to find self-consistent solutions for the different physical processes in the model.
A first step in this direction will be the implementation of a multi-grid radiative transfer solver in \Magritte{}.

Currently, mesh generation is a separate pre-processing step to the radiative transfer simulation.
However, in order to effectively leverage the methods presented here, for instance in coupled radiation-hydrodynamics simulations, mesh generation should happen on-the-fly and preferably in a way that is aware of the evolution of the mesh.

\section{Conclusion}
\label{sec:conclusion}
Radiative transfer models are a crucial but very computationally demanding component of almost all astrophysical and cosmological simulations.
In this paper, we demonstrate how the choice of both the directional and spatial discretization scheme can help alleviate the computational cost.
First, we presented the improved adaptive ray-tracing scheme implemented in the 3D radiative transfer library \Magritte{}, that uses an adaptive hierarchical scheme to discretize directions based on \textsc{HEALPix} \citep{Gorski2005}.
Second, we demonstrated how the free and open-source software library \textsc{Gmsh} \citep{Geuzaine2009} can be used to generate sparse meshes, even for morphologically complex models, that are ideally suited for radiative transfer simulations.
Furthermore, we proposed two simple algorithms, one for analytically and one for numerically defined models, which can extract a desired mesh element size distribution from a model that will result in a sparse mesh, while preserving a proper sampling of key model features.
Since typically the output of hydrodynamics models is used as input for radiative transfer simulations, we applied these algorithms to snapshots of several hydrodynamic models and showed that the number of elements can be reduced by an order of magnitude, without a significant loss of accuracy in the computation of the radiation field.
As a result, the radiation field on the reduced meshes can be computed more than an order of magnitude faster.
This reduced computational and memory cost can either be used to speedup the computation or it can be invested to increase the accuracy by refining the mesh in critical locations.
The examples included both models based on an hierarchical octree mesh resulting from adaptive mesh refinements, as well as smoothed-particle hydrodynamics data.
We conclude that carefully constructing an appropriate directional and spatial discretization scheme, using the methods described above, can significantly decrease the computational cost of radiative transfer simulations and make feasible simulations that would otherwise be intractable.

\section*{Acknowledgements}
We would like to thank Steven Dargaville and Tom Deakin for organising the first meeting on Boltzmann transport applications at Imperial College London, bringing together researchers working on Boltzmann transport from a broad range of disciplines, ranging from nuclear engineering, to medical imaging, and astronomy.
We would like to thank Roald Frederickx for stimulating discussions on ray-tracing and mesh construction in computer science.
FDC is supported by the EPSRC iCASE studentship programme, Intel Corporation and Cray Inc. FDC, JB, WH, and LD acknowledge support from the ERC consolidator grant 646758 AEROSOL.
This work was performed using the Cambridge Service for Data Driven Discovery (CSD3), part of which is operated by the University of Cambridge Research Computing on behalf of the STFC DiRAC HPC Facility (\href{https://dirac.ac.uk}{www.dirac.ac.uk}).
The DiRAC component of CSD3 was funded by BEIS capital funding via STFC capital grants ST/P002307/1 and ST/R002452/1 and STFC operations grant ST/R00689X/1. DiRAC is part of the National e-Infrastructure.

\section*{Data availability}
The models underlying this article and the corresponding scripts to process them can be found in the online repository \href{https://github.com/Magritte-code/Examples}{github.com/Magritte-code/Examples}.



\bibliographystyle{mnras}
\bibliography{references}





\bsp	
\label{lastpage}
\end{document}